# Energy-level quantization in YBa$_2$Cu$_3$O$_{7-x}$ phase-slip nanowires

M. Lyatti[1,2*], M. A. Wolff[2], I. Gundareva[1,3], M. Kruth[4], S. Ferrari[2], R. E. Dunin-Borkowski[3], C. Schuck[2]

[1]Kotelnikov IRE RAS, 125009 Moscow, Russia

[2]Institute of Physics, University of Münster, 48149 Münster, Germany

[3]PGI-5, Forschungszentrum Jülich, 52425 Jülich, Germany

[4]ER-C 2, Forschungszentrum Jülich, 52425 Jülich, Germany

**Significant progress has been made in the development of superconducting quantum circuits, however new quantum devices that have longer decoherence times at higher temperatures are urgently required for quantum technologies. Superconducting nanowires with quantum phase slips are promising candidates for use in novel devices that operate on quantum principles. Here, we demonstrate ultra-thin YBa$_2$Cu$_3$O$_{7-x}$ nanowires with phase-slip dynamics and study their switching-current statistics at temperatures below 20 K. We apply theoretical models that were developed for Josephson junctions and show that our results provide strong evidence for energy-level quantization in the nanowires. The crossover temperature to the quantum regime is 12-13 K, while the lifetime in the excited state exceeds 20 ms at 5.4 K. Both values are at least one order of magnitude higher than those in conventional Josephson junctions based on low-temperature superconductors. We also show how the absorption of a single photon changes the phase-slip and quantum state of a nanowire, which is important for the development of single-photon detectors with high operating temperature and superior temporal resolution. Our findings pave the way for a new class of superconducting nanowire devices for quantum sensing and computing.**

Superconducting quantum circuits are based on electrical (LC) oscillators, in which the Josephson effect contributes the nonlinearity that is required for selective access to quantum levels[1]. Historically, tunnel Josephson Junctions (JJ) have played a major role in studies of macroscopic quantum phenomena and, nowadays, most superconducting quantum circuits are based on them. However, the Josephson effect also occurs in structures with a non-tunneling conductivity, which together with tunnel JJs form a class of superconducting weak links[2]. Nanowires with quantum phase slips are particularly interesting superconducting weak links with direct conductivity as they can be used in superconducting quantum circuits[3-6]. The potential of these nanowires lies in their long-lived excited states, which result from their low sensitivity to charge noise and critical current noise[3]. The electrodynamics of superconducting nanowires with quantum phase slips are not well understood, but are likely to be governed by principles similar to those for single JJs. Both systems can be described by the resistively and capacitively shunted junction (RCSJ) model[7-9]. Here, we use the term phase-slip nanowire (PSN) to refer to a superconducting nanowire that has a finite critical current, with the resistive state occurring due to phase slippage.

* matvey_l@mail.ru

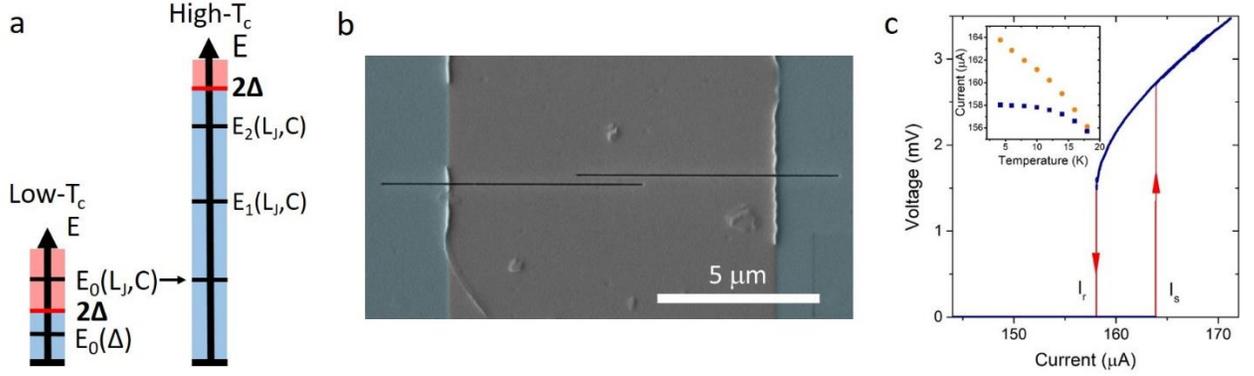

**Figure 1| a,** Energy diagrams for low-$T_c$ and high-$T_c$ nanowires**.** Energies below and above 2Δ are shown in blue and red, respectively. **b,** Scanning electron micrograph of an YBCO nanowire shaped by two FIB cuts across a microbridge. The YBCO film, STO substrate and FIB cuts are shown in gray, blue and black, respectively. **c,** IV curve of a 55-nm-wide YBCO nanowire at 4.2 K. Inset: temperature dependences of the average switching $I_s$ (orange circles) and retrapping $I_r$ (blue squares) currents.

A tunnel JJ exhibits two macroscopic quantum phenomena that are important for quantum circuits: macroscopic quantum tunneling (MQT)[10] and energy-level quantization (ELQ)[11]. MQT has been demonstrated for PSNs fabricated from different low-temperature (low-$T_c$) superconductors[12-14]. However, there is a significant difference between a JJ and a low-$T_c$ PSN, because the physical mechanisms that determine the frequency of plasma oscillations have a very different nature. The zero-bias plasma frequency of a JJ $\omega_{p0} = (2eI_c/C\hbar)^{1/2}$ (referred to as a Josephson plasma frequency) is given by the resonant frequency of an $L_JC$ circuit consisting of a Josephson inductance $L_J = \hbar/2eI_c cos\varphi$ and a junction capacitance $C$ [15], where $I_c$ is the junction critical current, $e$ is the electron charge, $\hbar$ is the reduced Planck constant and $\varphi$ is the phase difference across the junction. When compared with tunnel JJs, the nanowires typically have very small intrinsic capacitances. In a low-$T_c$ nanowire, the energy of Josephson plasma oscillations is much higher than the superconducting energy gap 2Δ, which makes such oscillations impossible, as shown in Figure 1a. In a pioneering study[14], Giordano proposed that the plasma frequency of low-$T_c$ PSNs is limited by another physical mechanism and scales with Δ. This was experimentally confirmed by measurements of the crossover temperature between MQT and thermal activation (TA) escape mechanisms for different low-$T_c$ PSNs[12,13]. As a result of the very high plasma frequency, it is therefore unlikely that more than one energy level exists in a low-$T_c$ PSN. In contrast, in high-temperature (high-$T_c$) superconductors the superconducting energy gap is much larger and Josephson plasma oscillations are allowed, as shown in Figure 1a. In addition to having a large superconducting energy gap of $\Delta$ = 25–30 meV [16], YBa$_2$Cu$_3$O$_{7-x}$ (YBCO) nanowires meet the requirements for the "ideal" Josephson effect[2]: they exhibit Josephson behavior [17] and show a single-valued sine-like current-phase relationship, even at temperatures close to zero[18]. Here, we show that YBCO nanowires are promising candidates for realizing superconducting quantum circuits. Our measurements of switching-current statistics for ultra-thin YBCO nanowires with phase-slip dynamics[19] provide clear evidence of ELQ in the nanowires.

**Prospect of energy-level quantization in YBCO phase-slip nanowires**

We performed electrical transport measurements on 2-μm-long 8.2-nm-thick YBCO nanowires with widths below 160 nm on a (100) SrTiO$_3$ (STO) substrate. Figure 1b shows a scanning electron micrograph of a representative nanowire patterned using focused ion beam (FIB) milling across a 10-μm-wide microbridge. All of the nanowires showed current-voltage (*IV*) curves that were characteristic of phase slippage. Based on the linear dependence of switching current on nominal nanowire width, we determined the effective nanowire width and thickness. Significantly, the switching-current statistics of nanowires with effective widths of below 100 nm cannot be explained using models developed for low-$T_c$ nanowires. For a nanowire with an effective thickness and width of 4.3 and 55 nm, respectively, we observe current-voltage characteristics that show direct voltage switching from the superconducting to the resistive state and large current hysteresis over an extended temperature range of up to 18–20 K, as shown in Figure 1c.

The zero-bias plasma frequency of the 55-nm-wide nanowire is calculated as $\omega_{p0}/2\pi \approx 1.6$ THz for a nanowire capacitance of $C$ = 4.8 fF (see the Methods section) and on the assumption that the switching current is close to the critical current. The energy of the plasma oscillations $\omega_{p0}\hbar$ = 6.6 meV is significantly lower than $2\Delta_0$ = 50-60 meV. Therefore, several quantized energy levels can exist in the nanowire. The detection of a macroscopic quantum tunneling event, which can be considered as particle escape from a well of a tilted washboard potential, depends on the quality factor $Q_n$ of the JJ/PSN in the resistive state. For $Q_n > 0.8382$ the particle continues to move along a tilted washboard potential after escape, even when the washboard potential has local minima (referred to as a running state)[20]. This running state is observed as a jump from the superconducting to the resistive state and is straightforward to detect. Here, we use the motion of a particle in a tilted washboard potential as a mechanical analog of the RCSJ model. We calculate the quality factor of the nanowire in the resistive state as $Q_n = \omega_{ps}R_nC_{ps} = 2\omega_{ps}R_nC_S\Lambda_Q \approx 1$, where $R_n$ = 120 Ω is the phase-slip line resistance, $\omega_{ps} = 2eV_s/\hbar$ is the phase-slip oscillation frequency, $V_s$ = 2.7 mV is the voltage switching amplitude, and $C_{ps}$ = 1 fF is the capacitance that shunts the phase-slip oscillations.

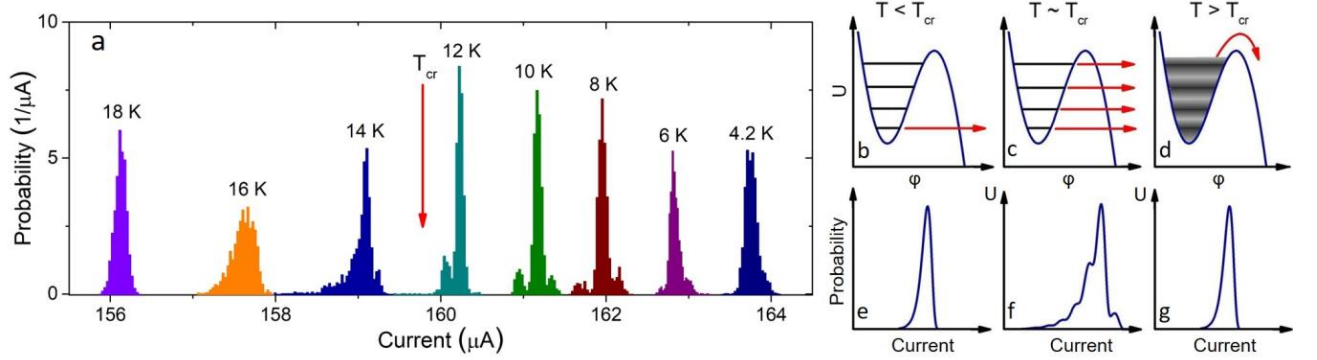

**Figure 2| a,** Switching-current distributions for a 55-nm-wide YBCO nanowire. **b-g,** Washboard potential with quantum states (b-d) and corresponding switching current distributions (e-f) measured with a "non-adiabatic" sweep rate at temperatures below, close to and above the crossover temperature between the MQT and TA regimes.

In order to calculate the capacitance $C_{ps}$, we take into account that the phase-slip oscillations take place in a $2\Lambda_Q$-long section of the nanowire[9], where $\Lambda_Q = 200$ nm is the charge imbalance distance[19]. Furthermore, we expect a crossover between TA and MQT escape mechanisms at a temperature of $T_{cr} \approx \omega_{p0}\hbar/2\pi k = 12.4$ K, where $k$ is the Boltzmann constant[15]. Based on these calculations, we conclude that the energy levels in our YBCO phase-slip nanowires must be quantized and can be revealed using switching-current measurements.

**Experimental results and discussion**

We measured switching-current statistics for a 55-nm-wide 4.3-nm-thick YBCO nanowire, both under equilibrium conditions and under illumination with 77 K black body radiation (BBR) as well as visible light, inducing non-equilibrium state of the wire. In order to reach thermal equilibrium of the nanowire with external radiation, we kept the radiation shield surrounding the nanowire and the nanowire itself at the same temperature. We recorded 1500 IV curves with a current sweep rate of $dI/dt = 0.55$ mA/s for each of eight distinct temperatures in the 4.2-18 K temperature range, determined the switching current values and extracted the switching-current distributions (SCDs), which are shown in Figure 2a. At the highest temperature of 18 K, the SCD shows a single peak at a switching current of 156.12 μA. As the temperature is decreased to 16-14 K, the SCD peak shifts to higher switching currents, broadens and develops an asymmetry with a fine structure of closely-spaced peaks with spacings of 74±20 nA and 116±31 nA, respectively (enlarged SCDs are available in the Supplementary Information). Below 14 K, the SCDs show new peaks on either side of the main peak. This three-peaked structure is most pronounced close to 10 K and gradually disappears at 6-4.2 K.

Within the framework of the RCSJ model, the switching and retrapping processes are affected by external noise in a similar way. We measure retrapping current distributions (RCDs) simultaneously with the SCD and find only a single peak with standard deviation $\sigma_r$ = 62.8±3.6 nA for all temperatures, indicating that the SCD transformation is caused by intrinsic nanowire dynamics rather than by external noise (Supplementary Information).

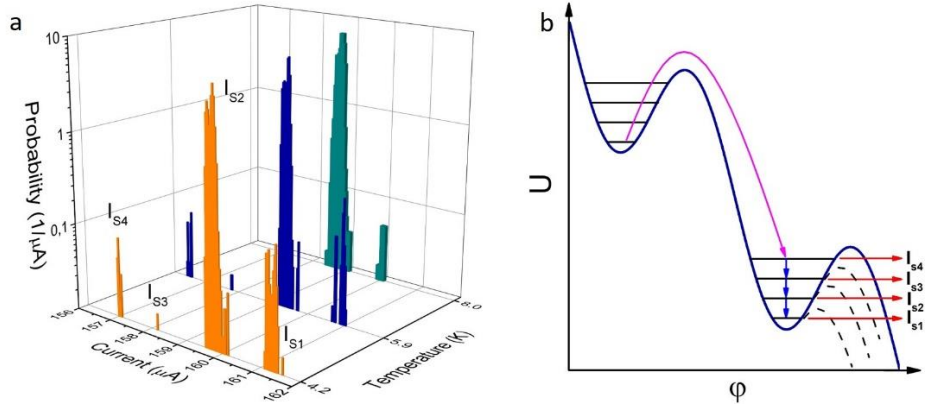

**Figure 3| a,** Switching-current distributions of the 55 nm-wide YBCO nanowire affected by low-frequency noise. **b,** Noise-induced escape from a local minimum of the tilted washboard potential (magenta arrow) and subsequent escapes into the resistive state by tunneling from the different energy levels (red arrows). Blue arrows correspond to transitions from the upper to the lower energy level. Dashed lines are the washboard potentials at $I = I_{s1}\text{-}I_{s3}$.

The broadening of the SCD at temperatures slightly above the calculated crossover temperature $T_{cr}$ = 12.4 K is consistent with the temperature dependence of SCDs for the tunnel JJ measured using a "non-adiabatic" current sweep rate[21]. The "non-adiabatic" transition between the TA and MQT regimes is shown schematically in Figures 2b-g. At temperatures well below $T_{cr}$, only the lowest (ground) energy level is populated (Figure 2b). Escape is dominated by MQT from the ground state energy level and the corresponding SCD is single-peaked and narrow (Figure 2e). When the temperature is close to $T_{cr}$, higher energy levels become populated as a result of thermal fluctuations (Figure 2c), according to a Boltzmann distribution. If the current sweep rate is high, so that the upper energy level cannot be refilled by thermal fluctuations, then escape from the lower energy levels becomes possible and the nanowire can switch into a resistive state over a wider current range[21]. The corresponding SCD (Figure 2f) is broad and has a fine structure of closely-spaced peaks caused by escape from different energy levels. At temperatures well above the crossover temperature, the energy levels are broadened and their superposition forms a continuous energy band, as shown in Figure 2d. The nanowire switches into the resistive state *via* thermal activation from the upper edge of this energy band, resulting in a single-peaked and narrow SCD (Figure 2g). "Non-adiabatic" broadening of the SCD can be observed experimentally when the current sweep rate $dI/dt > I_c/200R_{qp}C$, where $R_{qp}$ is the quasiparticle resistance[21]. By using this expression, we estimate a quasiparticle resistance $R_{qp} > 3\cdot10^{11}$ $\Omega$ and find the quality factor in the superconducting state $Q_s \approx \omega_{p0}R_{qp}C > 1.5\cdot10^{10}$ and the lifetime in the excited state $\tau \approx Q_s/\omega_{p0} > 1.5$ msec at $T \approx T_{cr}$. We attribute the presence of smaller peaks on both sides of the main peak in the SCD at $T < T_{cr}$ to interaction of the nanowire plasma oscillations with an external resonant system, *i.e.*, with geometric resonances of the 10-µm-long microbridge or the bow-tie antenna of our device (see Methods), with resonant frequencies of 1.2 THz and 40 GHz, respectively.

In order to probe the energy level structure in the PSN under equilibrium conditions, we applied external noise to the nanowire *via* the cables connecting the Dewar insert to the measurement equipment. We found that external noise has a significant effect on the SCDs, as shown in Figure 3a for three different temperatures at a current sweep rate $dI/dt = 2.75$ mA/s. At 4.2 K, the SCD consists of four nearly-equally-spaced peaks, which are marked $I_{s1}$- $I_{s4}$ in Figure 3a. At 5.9 K, the spacing between the peaks decreases slightly. At 8 K, only two peaks remain in the SCD. The noise-affected RCD shows only a single peak with $\sigma_r = 51 \pm 3$ nA (Supplementary Information), indicating that the applied low frequency noise does not cause the multiple peak structure in the SCD, but acts as a trigger for another physical mechanism, which is discussed below.

The nearly-evenly-spaced peaks $I_{s1}$-$I_{s4}$ in Figure 3a are signatures of tunneling from different energy levels, as shown in Figure 3b. If the damping is moderate, *i.e.*, if the quality factor is close to unity, then a particle that escaped from the potential well due to external noise activation (magenta arrow) can be trapped in the upper energy level of the lower potential well. This retrapping process in a PSN is similar to that of thermal or quantum phase diffusion in underdamped tunnel-JJs[22,23]. When a particle is trapped in the upper energy level, it can decay to lower energy levels in the same well (blue arrows) or escape from the potential well by tunneling (red arrows), resulting in the presence of multiple peaks in the SCD. Within the framework of this model, the $I_{s1}$-$I_{s4}$ values are given by a system of equations $\Delta U(I_{si}) = \Delta U_{tun} + E_n(I_{si})$ (1), where $i$ is an integer and $\Delta U(I_{si})$ and $E_n(I_{si})$ are the energy barrier height and the energy of the populated energy level corresponding to escape at current $I_{si}$, respectively. Here, we assume that the particle tunnels through the barrier when the barrier height for the populated energy level is decreased to $\Delta U_{tun}$. We approximate the nanowire by a harmonic quantum oscillator with energy levels $E_n(I_{si}) = \omega_p(I_{si})\hbar(n+1/2)$, where $\omega_p$ is the current-dependent plasma frequency. We also use the approximate expressions for barrier height $\Delta U(I) = (hI_c/2\pi e)[(1-(I/I_c)^2)^{1/2}-(I/I_c)\arccos(I/I_c)]$ [24] and plasma frequency $\omega_p(I) = \omega_{p0}[1-(I/I_c)^2]^{1/4}$ [15]. By solving this system of equations, we obtain $\omega_{p0}/2\pi = 744$ GHz. We analyze the stability of the solution under small perturbations of the initial parameters and find that the real-valued solution disappears when the $I_{si}$ values are varied by more than 100 nA from their initial values. Hence, we conclude that the observed peaks $I_{s1}$-$I_{s4}$ can indeed be assigned to energy levels in a tilted washboard potential. We note that the calculated zero-bias plasma frequency is only approximately half of our previous estimate based on the nanowire capacitance and the crossover temperature. Using the numerically simulated eigenvalues $E_n/\omega_p\hbar$ of the Josephson junction in a cubic approximation of the washboard potential, which take values of 0.5, 1.45, and 2.37 for $n = 0$, 1 and 2 [24], we obtain the $\omega_{p0}/2\pi$ value as 1.5 THz, which is close to our theoretical estimate. The reduced number of peaks in the SCD at 8 K can be attributed to a decrease in energy level lifetime with increasing temperature when transitions to the lower energy level become more probable than tunneling through the barrier.

In order to probe the energy level structure of the nanowire using external radiation, we illuminated the device with 77 K BBR, which has a broad continuous spectrum peaked at a frequency of 8 THz and can populate energy levels up to $2\Delta$, resulting in a non-equilibrium state of the nanowire. We studied the resulting non-equilibrium nanowire state in the 5.4-20.1 K temperature range using three current sweep rates of 0.055, 0.55, and 2.75 mA/s. The resulting SCDs (orange bars) are shown in Figure 4a-d. Well above the crossover temperature, the SCD does not show any sign of quantized energy levels (Figure 4a), similar to experiments with the nanowire under equilibrium conditions. When the temperature is close to the crossover temperature (Figure 4b), the SCD has a single peak at the lowest current sweep rate $dI/dt$ = 0.055 mA/s, which broadens towards higher currents with increasing current sweep rate, eventually transforming into a distribution with two peaks at $dI/dt$ = 2.75 mA/s. This transformation of the SCD with current sweep rate reflects a transition from "adiabatic" to "non-adiabatic" measurements, as lower energy levels become accessible. Below the crossover temperature, at 10.9 K, we observe a strong dependence of the spectral weight of the peak $I_{s2}$ on the current sweep rate, as shown in Figure 4c. At even lower temperature (5.4 K in Figure 4d), the spectral weight dependence of the $I_{s2}$ peak with current sweep rate is less pronounced and the $I_{s1}$ peak is hardly observable.

We interpret the difference in spectral weight between the peaks at $I_{s1}$ and $I_{s2}$ in terms of a population inversion resulting from the decay of higher-lying energy levels, similar to observations in superconducting quantum circuits based on tunnel JJs [25,26]. We identify the $I_{s1}$ and $I_{s2}$ peaks in Figure 4c-d with the ground and first exited energy levels, respectively. The effects of temperature and current sweep rate on the spectral weights of the $I_{s1}$ and $I_{s2}$ peaks can then be assessed using the tilted washboard potential model, as shown in Figure 4e. Tunneling from the ground, first and second excited energy levels occurs at currents $I_1$, $I_2$ and $I_3$, respectively. Population inversion in the steady state is possible for currents $I < I_3$ when three or more energy levels exist in the potential well. At $I_3 \leq I < I_2$, the number of the energy levels is reduced to two and the nanowire state evolves towards a steady state, with similar population probabilities for the ground and first excited states by spontaneous and stimulated transitions during the time $t_{32} = [I_3 - I_2]/(dI/dt) \approx [I_2 - I_1]/(dI/dt)$. The nanowire quantum state can be read out at $I = I_2$ or $I_1$. As the energy-level occupation probability depends strongly on temperature and current bias sweep rate, the spectral weight of the SCD peak at $I = I_2$ decreases with increasing temperature or with decreasing current sweep rate.

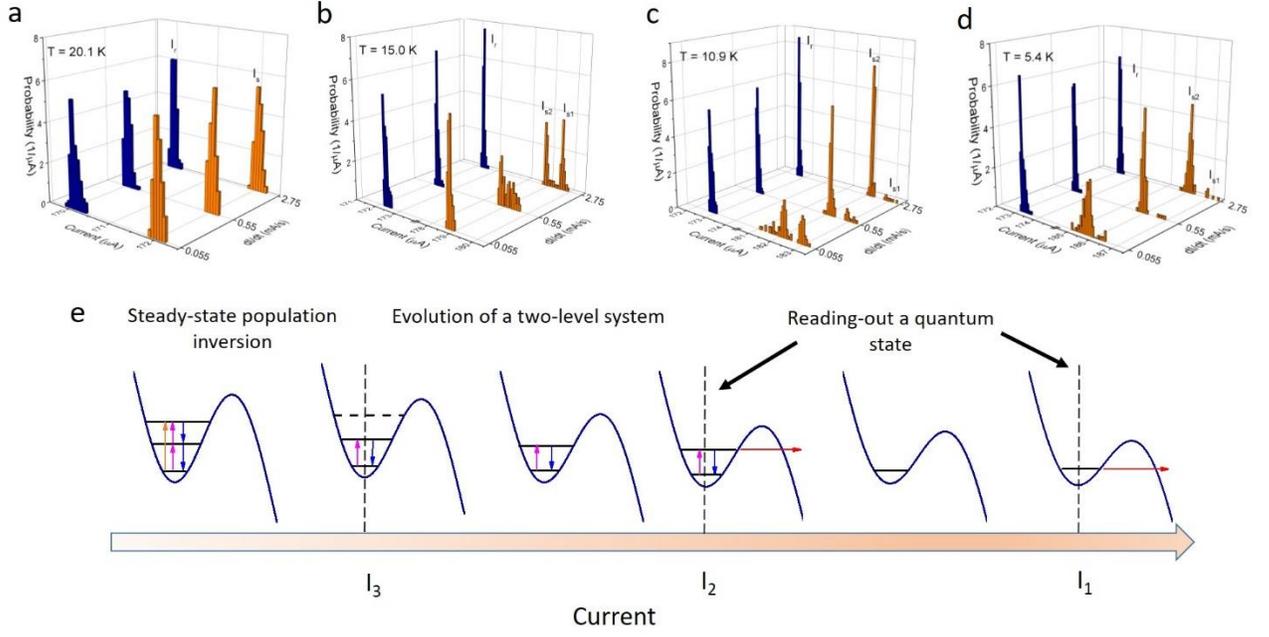

**Figure 4| a-d,** Switching- (orange) and retrapping-current (blue) distributions for a 55-nm-wide YBCO nanowire at different temperatures. **e,** Energy levels in the local minimum of the tilted washboard potential at different bias currents for the nanowire illuminated using 77 K blackbody radiation. Transitions between energy levels are shown using orange, magenta and blue arrows. Escape in the resistive state by tunneling is indicated by red arrows.

We calculate the lifetime in the exited state by associating the spectral weight of the $I_{s2}$ peak with the occupation probability of the first excited energy level $P_2(t)$. A fit of the expression $A + B \cdot exp(-t/\tau_2)$ to the experimental data (Supplementary Information) yields reasonable fitting parameters of $A = 0.55$ and $B = 0.39$ and an excited state lifetime of $\tau_2 = 8.3$ ms. Assuming that the lifetime $\tau_2$ at 10.9 K is limited by quasiparticle losses and considering the nanowire as a lumped element with quasiparticle resistance $R_{qp} = R_n \cdot e^{\Delta/kT}$, where $R_n \approx 4200$ $\Omega$ is the normal-state nanowire resistance, we obtain a reasonable estimate of the superconducting energy gap in YBCO $\Delta(10.9K) = kTln(\tau_2/2\pi CR_n) \approx 17$ meV. The lifetime in the excited state $\tau_2$ at 5.4 K is, however, much longer than $t_{32} \approx 20$ ms, because we only observe a minimal dependence of the population probability $P_2(t)$ of the first exited level on the current sweep rate down to $dI/dt = 55$ μA/s. The RCDs show only a single peak for all temperatures and current sweep rates, as illustrated by the blue bars in Figure 4a-d. We find retrapping currents of 88.1±5.7, 67.9±4.4, and 56.8±5.0 nA at $dI/dt$ of 0.055, 0.55, and 2.75 mA/s, respectively and are able to reproduce similar $\sigma_r$ values with two different experimental setups, confirming that switching into the resistive state originates from internal nanowire dynamics and not external noise.

Additional illumination of the nanowire using optical radiation with an LED led to seemingly counterintuitive results: the number of switching events with higher switching current was observed to increase with increasing LED intensity. Representative SCDs measured at different intensities of a blue LED with 460 nm wavelength for a current sweep rate of $dI/dt = 0.55$ mA/s are shown in Figure 5a. When the LED is turned off, the nanowire is still subject to 77 K BBR and switches into the resistive state in current range II (see Figure 5a), which corresponds to tunneling from the first exited state of the nanowire. At an LED intensity of $I_{LED} = 0.7$ W/m$^2$, switching events occur not only in region II but also in region I of the current range in Figure 5a, which corresponds to tunneling from the ground energy state. As the LED intensity rises, the number of switching events in current region I increases and some switching events start to appear in the gap between current regions I and II, forming a peak at $I = 186.5$ µA when $I_{LED} = 8.3$ W/m$^2$.

The interaction of optical photons with a superconducting nanowire has been studied widely because of its practical importance for the development of superconducting nanowire single-photon detectors. Here, we use a refined hotspot model to analyze the effects of optical radiation on the YBCO nanowire[27]. An optical photon whose energy is much higher than the superconducting energy gap disrupts tens of Cooper pairs, resulting in the appearance of non-equilibrium quasiparticles. The absorbed photon induces a normal-state domain (hotspot) across the nanowire when the number of non-equilibrium quasiparticles reaches $N_q = n_s W d (\pi D \tau_{th})^{1/2}(1-I/I_c)$, where $n_s$ is the local density of paired electrons, $d$ is the nanowire thickness, $D$ is the quasiparticle diffusion coefficient, $\tau_{th}$ is the quasiparticle thermalization time and $I_c$ is the nanowire critical current[27]. If the entire photon energy $E_{ph}$ is transferred to the quasiparticles, their actual number is given by $N_q = E_{ph}/\Delta$. The boundary for hotspot appearance can then be calculated as $I_{HS} = I_c - [j_c E_{ph}/n_s \Delta (\pi D \tau_{th})^{1/2}]$, where $j_c$ is the critical current density. Photon absorption below and above $I_{HS}$ has qualitatively different consequences. For $I < I_{HS}$, the photon creates non-equilibrium quasiparticles, but the normal-state domain across the nanowire does not appear. For $I > I_{HS}$, photon absorption results in a hotspot across the nanowire, which leads to local collapse of the order parameter. The PSN evolves from this transient state towards a state with a phase-slip process, corresponding to switching of the nanowire from the superconducting to the resistive state, as described in Ref. [28].

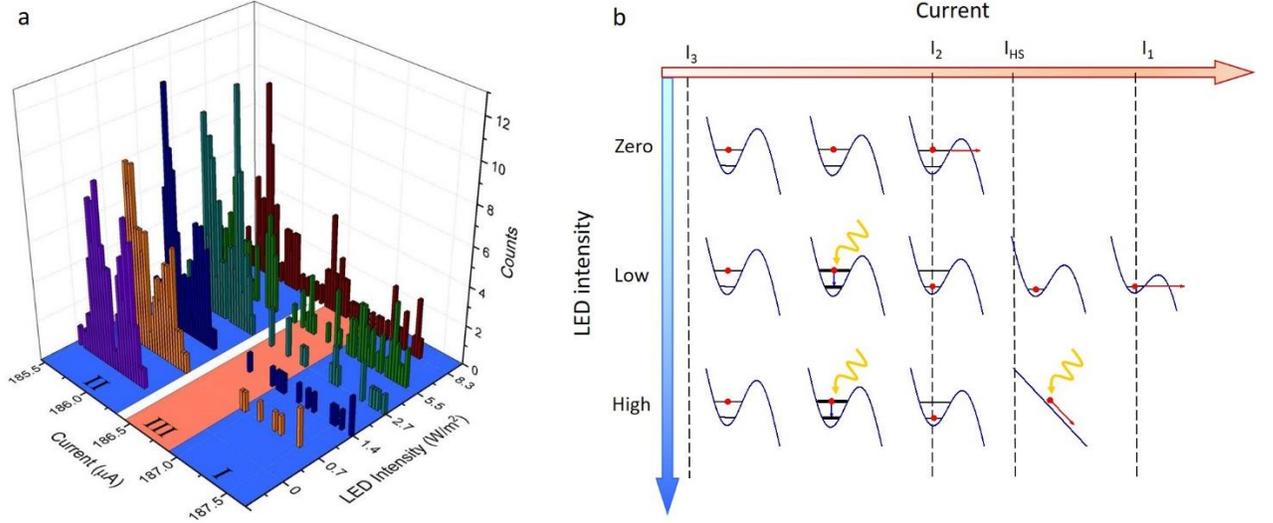

**Figure 5| a,** Switching-current distributions of a 55-nm-wide YBCO nanowire measured at 5.4 K for different LED intensities. Regions I and II (blue) correspond to ground and exited energy states of the nanowire. Region III (red) is the current region, in which switching can only occur by the hotspot effect. **b,** Illustration of the interaction of optical photons with the YBCO nanowire. The populated energy level is shown by a red dot. Photon absorption and escape events are indicated by wavy yellow and red arrows, respectively. Collapse of the order parameter after photon absorption is shown by an inclined blue line. The currents $I_1$, $I_2$ and $I_3$ are the same as in Figure 4e. The current $I_{HS}$ indicates the boundary of the hotspot regime.

For photons of wavelength 460 nm, we calculate $I_c$ - $I_{HS} = j_c E_{ph}/n_s\Delta(\pi D\tau_{th})^{1/2} = 1.1$-1.3 μA using $n_s = 1.1\cdot10^{27}$ m$^{-3}$ [29], $\Delta = 25$-30 meV, $D = 20$ cm$^2$/s [30], $\tau_{th} = 0.56$ ps [31], $E_{ph} = 2.7$ eV and the experimentally measured $j_c = 78.6$ MA/cm$^2$. In Figure 5a, we highlight this current region III in red, in which the nanowire can switch to the resistive state only by the 460 nm wavelength photons. We assume that the critical current in the refined hotspot model corresponds to the switching current of the nanowire in the ground energy state.

Figure 5b illustrates schematically the interaction of optical photons with a PSN that has quantized energy levels and is prepared in the excited state. When the LED is turned off, the nanowire switches to a resistive state at current $I_2$ due to tunneling from the first excited energy state. At a low LED intensity ($I_{LED} = 0.7$ W/m$^2$ in Figure 5a), the nanowire can absorb the photon before reaching current $I_2$. Since photon absorption takes place at $I < I_{HS}$, it creates a number of non-equilibrium quasiparticles, resulting in an increase in nanowire losses, which is shown in Figure 5b by a broadening of the energy levels and, hence, faster decay from the excited to the ground energy state. Switching into the resistive state then occurs at the higher current $I_1$ by tunneling from the ground energy level. Therefore, the nanowire quantum state can be changed by single-photon absorption.

When the radiation intensity is increased (corresponding to high LED intensity in Figure 5b and $I_{LED} = 1.4$-$5.5$ W/m$^2$ in Figure 5a), the nanowire can absorb a second photon at bias currents of $I > I_{HS}$, where hotspot conditions are met. As a result of the local collapse of the order parameter, the oscillating phase-slip process appears and the nanowire switches into the resistive state. Since the absorbed photon changes the number of phase-slip processes in the nanowire, *i.e.*, the phase-slip state of the nanowire, we refer to this process as the phase-slip mechanism for photon detection.

At very high radiation intensities, switching of the nanowire occurs predominantly at the boundary of the hotspot region, forming a peak in the SCD at $I = I_{HS}$. We observe this very high intensity regime at $I_{LED} = 8.3$ W/m$^2$ in Figure 5a, where a peak at the edge of region III is visible. We treat the observed interaction of the optical photons with the YBCO nanowire as a single-photon process because the position of the photon-induced SCD peak corresponds to the single-photon energy and the radiation power is much lower than that required for the two-photon process, taking into account the short quasiparticle recombination time in YBCO[30,32].

In this work, we have implicitly considered a nanowire made from YBCO, which has d$_{x^2-y^2}$-wave symmetry of the order parameter as a fully gapped superconductor. We find that a small ($N_q \leq 100$) number of non-equilibrium quasiparticles generated by an optical photon is significantly larger than the number of equilibrium quasiparticles in the nanowire, resulting in fast decay from the excited to the ground state. This behavior is expected for a fully gapped superconductor with an exponentially small number of equilibrium quasiparticles at a temperature well below the critical temperature. Deviations from d$_{x^2-y^2}$-wave symmetry in our YBCO nanowires can arise from size or doping effects, which have recently been observed in different cuprate superconductors[33,34].

**Conclusions**

We have fabricated sub-100-nm-wide YBCO nanowires with phase-slip dynamics and measured their switching-current statistics under equilibrium and non-equilibrium conditions. Our experimental data show energy-level quantization in YBCO phase-slip nanowires. The YBCO nanowires have a high crossover temperature between thermal activation and quantum regimes of 12–13 K and their lifetime in the excited state exceeds 20 ms at 5.4 K, which is at least one order of magnitude longer than in low-$T_c$ tunnel Josephson junctions[1]. We also show that the absorption of a single-photon changes the quantum and phase-slip states of YBCO nanowires. Our findings demonstrate that phase-slip YBCO nanowires are promising systems for quantum technology applications, including quantum sensing and computing.

**Acknowledgements**

M.L. thanks G. Catelani for valuable discussions. This work was partially supported by ER-C Project No. C-088. C.S. acknowledges financial support from the Ministry for Innovation, Science und Research (North Rhine-Westphalia).

**Author contributions**

M.L. and M.K. fabricated the nanowires. ML. and M.W. performed the measurements. S.F. and I.G. contributed to the experiments. M.L, I.G., R.D.B. and C.S. co-wrote the paper. All authors commented on the manuscript.

**Competing interests**

The authors declare no competing interests.

## Methods

**Nanowire fabrication.** YBCO nanowires were fabricated from an 8.2 nm (7 unit cell) thick YBCO film deposited on a $TiO_2$-terminated (100) STO substrate by dc sputtering at high (3.4 mbar) oxygen pressure. YBCO deposition followed a procedure that is described elsewhere [35]. 100-nm-thick Au contact pads were deposited *ex situ* using room temperature dc magnetron sputtering with a shadow mask. Following contact pad deposition, nanowires were fabricated in a two-stage process. In the first stage, 10-μm-wide 10-μm-long microbridges integrated with a bow-tie antenna and leads were patterned using optical UV contact lithography with a PMMA resist and Ar ion beam etching. In the second stage, 2-μm-long nanowires aligned along STO crystallographic axes were fabricated across the microbridges with two cuts made with focused ion beam (FIB) milling using a Au/PMMA protective layer. A sketch of the device is shown in Extended Data Fig. 1a. More details about the patterning process can be found in Ref. [19].

**Experimental setups.** We performed the measurements using two experimental setups. The first experimental setup was based on a liquid helium storage Dewar insert filled with He exchange gas, in which the nanowire and surrounding radiation shield had the same temperature. The second setup was based on an HLD-5 liquid helium cryostat (Infrared Laboratories, Inc.). The sample was placed on a sample holder mounted on the 4 K stage of the cryostat and shielded by a radiation shield with a quartz window. The sample was illuminated through the window using continuous optical radiation emitted by LEDs and 77 K blackbody radiation from a 77 K radiation shield. The LEDs were placed in front of the window at a distance of 2 cm from the substrate and cooled to 77 K. The sample holder temperature was maintained to an accuracy of ±5 mK at 4.2 K and ±20 mK at 20 K for the Dewar-insert-based setup and ±10 mK for the cryostat-based setup over the whole temperature range. The wiring inside the Dewar insert was made with twisted-pair cables and had a bandwidth of 3.9 MHz. The nanowires inside the cryostat were connected to room temperature measuring equipment using high-frequency SMA coax cables and a 10 GHz probe. We used battery-operated low-noise analog electronics with a 100 kHz frequency bandwidth to sweep the bias current and amplify the voltage across the nanowire. The root-mean-square noise of the current source was 1 nA. The output signals of the analog electronics, proportional to the current and voltage across the nanowire, were digitized using a simultaneous 16-bit data acquisition board DT9832 (Data Translation). All electrical connections, apart from the HF coax cables, were filtered using low-frequency feedthrough filters. Cables between the analog electronics and cryogenic units were as short as possible to eliminate electromagnetic interference. The frequency spectrum of the output voltage signal was controlled before the measurement to ensure that no low-frequency external noise was present in the measurement system.

**Switching-current measurement.** In order to measure the switching current- and retrapping-current statistics, the bias current through the nanowire was ramped linearly up and down over the range $(0.85–1.05)I_s$ and $IV$ curves were recorded with $2 \cdot 10^4$ points per curve. The spacing between independent current points of 15.3 nA was determined by the 16-bit resolution of the data acquisition board. The switching and retrapping currents were determined by post-processing of the recorded data. Standard deviations of the retrapping current were computed in the standard way.

**Nanowire capacitance calculation.** The nanowire layout shown schematically in Extended Data Fig. 1b is similar to that of a coplanar waveguide with ground. As a result of this similarity, we used a coplanar waveguide calculator[10] to calculate the nanowire capacitance. The following parameters were used for the capacitance calculations: spacing $S = 200$ nm, substrate thickness $H = 1$ mm, and dielectric index of the STO substrate $\varepsilon = 300$. The spacing $S$ consisted of the 60-nm-wide FIB cut and two 70-nm-wide damaged insulating regions of YBCO on the sides of the FIB cut [19]. By using the coplanar waveguide calculator [36], we obtained a nanowire specific capacitance per unit length of $C_S = 2.4$ fF/μm and a total capacitance of the 2-μm-long nanowire of C = 4.8 fF.

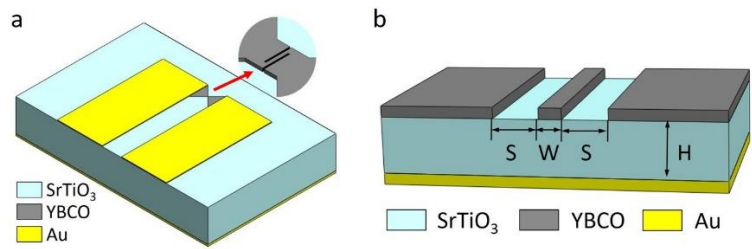

**Extended Data Figure 1| a,** Sketch of the whole nanowire device. **b,** Nanowire layout for capacitance calculations

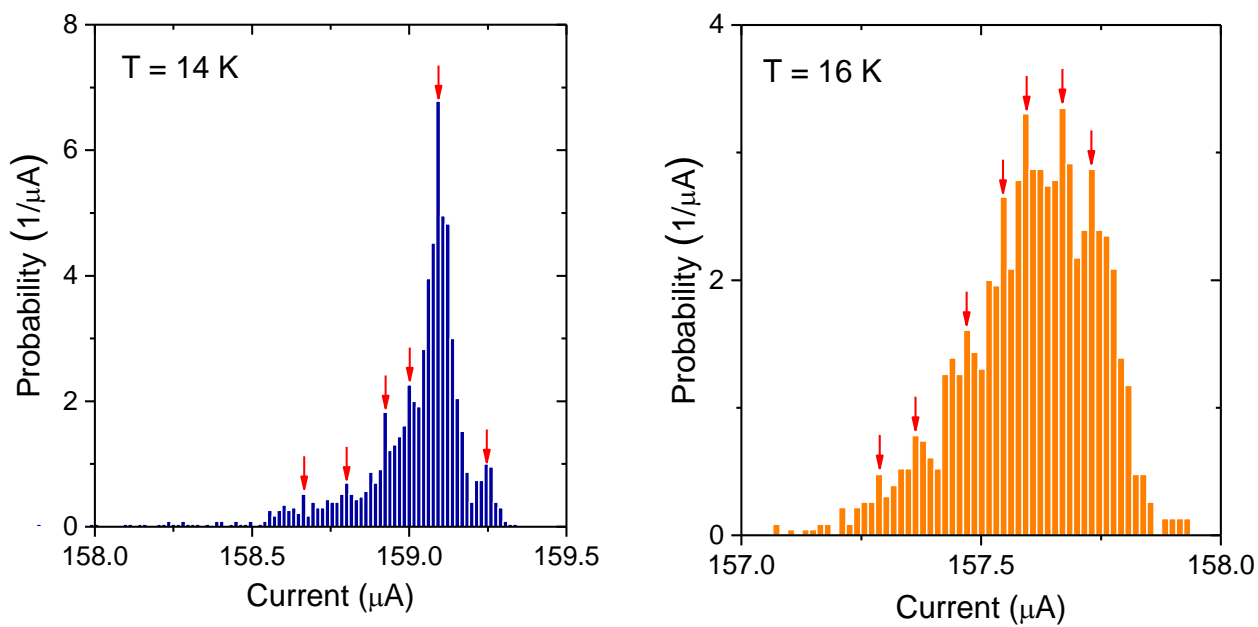

**Supplementary Fig. S1|** Switching-current distributions for a 55-nm-wide YBCO nanowire measured under equilibrium conditions at temperatures of 14 and 16 K. Red arrows indicate the bias current value at which the height of the energy level coincides with the barrier height.

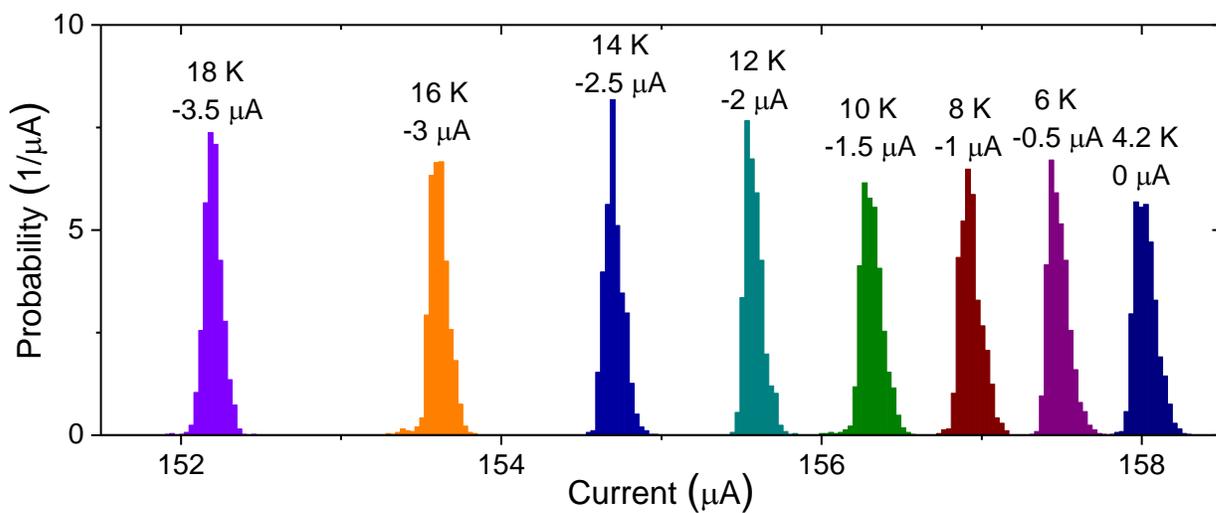

**Supplementary Fig. S2|** Retrapping-current distributions for a 55-nm-wide YBCO nanowire measured under equilibrium conditions. The numbers above the peaks are the nanowire temperature and the shift along the current axis.

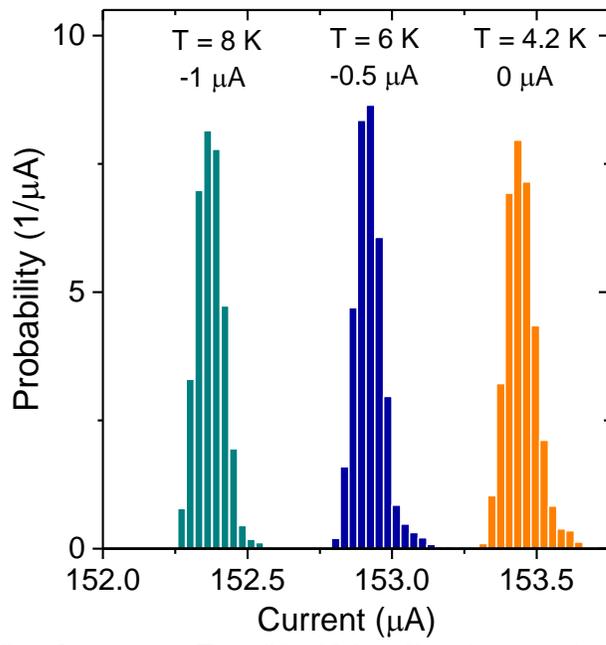

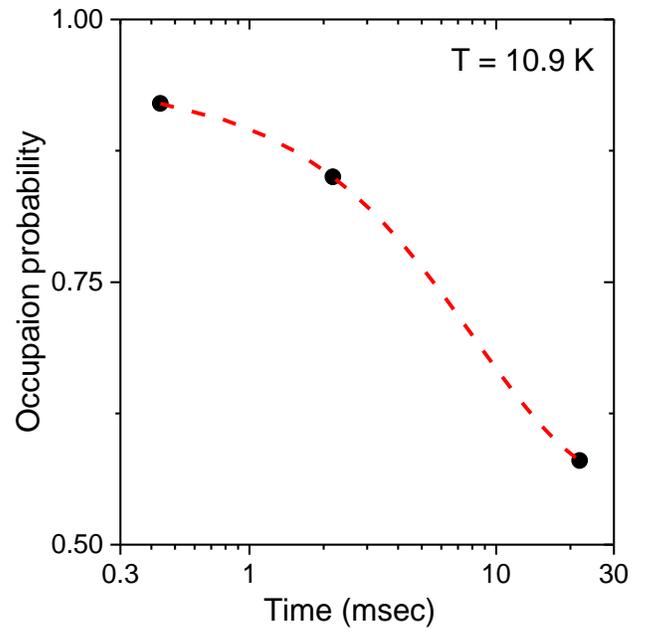

**Supplementary Fig. S3|** Noise-affected retrapping current distributions for a 55-nm-wide YBCO nanowire. The numbers above the peaks are the nanowire temperature and the shift along the current axis.

**Supplementary Fig. S4|** Time dependence of the occupation probability of the excited state for a 55-nm-wide YBCO nanowire at 10.9 K (black dots). The dashed line is an exponential decay fit to the experimental data.